\documentstyle[amssymb,12pt]{amsart}
\newcommand{\Z}{{\Bbb Z}}

\newcommand{\C}{{\Bbb C}}

\newcommand{\Ref}[1]{{\rm{(}\ref{#1}\rm{)}}}
\newcommand{\bean}{\begin{eqnarray}}
\newcommand{\eean}{\end{eqnarray}}
\newcommand{\be}{\begin{displaymath}}
\newcommand{\ee}{\end{displaymath}}
\newcommand{\bea}{\begin{eqnarray*}}
\newcommand{\eea}{\end{eqnarray*}}
\newcommand{\g}{{{\frak g}\,}}

\newcommand{\h}{{{\frak h\,}}}

\newcommand{\ad}{{\mbox{ad}}}

\newcommand{\noi}{\noindent}

\newtheorem%
{thm}{Theorem}
\newtheorem%
{proposition}[thm]{Proposition}
\newtheorem%
{lemma}[thm]{Lemma}
\newtheorem%
{lemmadef}[thm]{Lemma-Definition}
\newtheorem%
{corollary}[thm]{Corollary}
\newcommand{\End}{{\mbox{End}}}

\newcommand{\sign}{{\mbox{sign}}}
\title[Elliptic KZB equations]
{Integral representation of solutions of the elliptic
Knizhnik--Zamolodchikov--Bernard equations}
\thanks{Supported in part
by NSF grants DMS-9400841 and DMS-9203929.}
\author{Giovanni Felder \and Alexander Varchenko}
\address{
Department of Mathematics\\ Phillips Hall CB 3250\\
University of North Carolina at Chapel Hill\\
Chapel Hill, NC 27599-3250, USA%
}
\email{felder@@math.unc.edu, varchenko@@math.unc.edu}
\date{February 1995}
\begin{document}
\maketitle
\begin{abstract} We give an integral representation of
solutions of the elliptic {Kni\-zhnik--Za\-mo\-lo\-dchi\-kov--Ber\-nard}
equations for arbitrary simple Lie algebras.
If the level is a positive integer,
we obtain formulas for conformal blocks of the WZW model on
a torus. The
asymptotics of our solutions at critical level gives
eigenfunctions of Euler--Calogero--Moser integrable
$N$-body systems. As a by-product, we obtain some
remarkable integral identities involving classical
theta functions.
\end{abstract}
\medskip
\centerline{To appear in International Mathematics Research Notices}
\medskip
\section{Introduction}
The subject of this note is the set of
Kni\-zhnik--Za\-mo\-lo\-dchi\-kov--Ber\-nard
(KZB) equations,
obtained by Bernard \cite{B1,B2}
as a generalization of the KZ equations.

We consider here the case of elliptic curves
with marked points, in the more general context
of complex level. Then the KZB equations are
the equations for horizontal sections of an infinite
rank holomorphic vector bundle. If the level is
a positive integer, this vector bundle
has a finite rank subbundle preserved by the
connection, which is
relevant to conformal field theory.

In Section \ref{kzbe}, we define the KZB equations.
In Section \ref{kzbc} we interpret these
equations as the horizontality condition for a
connection on a holomorphic vector bundle,
and give (Section \ref{wgr}) an a priori regularity theorem for
Weyl antiinvariant meromorphic solutions.

We then give an integral representation of solutions of the KZB
equation. The integration cycles have coefficients in a local system
of infinite rank which can be viewed as the sheaf of local solutions
of an Abelian version of the KZB equation, see Section \ref{tls}.  In
Section \ref{tir}, the integrand is given in terms of ``elliptic
logarithmic forms'' by essentially the same combinatorial
formulas  as in the case of the Riemann sphere.

In the last section of this paper we give three applications: In the
case of conformal field theory, the average over the Weyl group of our
solutions belongs to the subbundle of conformal blocks (Theorem
\ref{cft}). At the critical level, we obtain, following
Etingof and Kirillov, Bethe ansatz eigenfunctions
for quantum $N$-body systems (Theorem \ref{lame}), generalizing the
work of Hermite on Lam\'e's equation.  And in special cases, where the
KZB equations can be solved by other means, we obtain integral
identities involving classical theta functions, see Theorem
\ref{intid}.

We restrict ourselves in this note to the case
of simple Lie algebras for clarity of exposition.
However, the proper context for our result
is the general setting of Kac--Moody Lie algebras
with symmetrizable Cartan matrix, as in \cite{SV}.
Also, the KZB connection can be interpreted
geometrically as a Gauss--Manin connection. These
aspects will be discussed elsewhere, along with proofs
of the results announced here.

\section{ The Knizhnik--Zamolodchikov--Bernard equation}\label{kzbe}
Let $\g$ be a simple complex Lie algebra.  Fix a Cartan subalgebra $\h$
and let
$\g=\h\oplus\Sigma_{\alpha\in\Delta}\g_\alpha$ be the corresponding root
space decomposition.  We identify $\h$ with
its dual space using the invariant bilinear form $(\ ,\ )$ on $\g$,
which is normalized in such a way that $(\alpha,\alpha)=2$, for long
roots $\alpha$.  The symmetric invariant tensor $C\in\g\otimes\g$ dual
to $(\ ,\ )$ has then a decomposition
$C_0+\Sigma_{\alpha\in\Delta}C_\alpha$, with $C_0\in\h\otimes\h$ and
$C_\alpha\in\g_\alpha\otimes\g_{-\alpha}$.

Let $\Lambda_1,\dots,
\Lambda_n\in\h^*$ be dominant integral weights, and $V_1$,
\dots, $V_n$ be the corresponding irreducible
highest weight $\g$-modules.  The KZB equations are equations for a
function $u(z_1,\dots,z_n,\tau,\lambda)$ with values in the weight
zero subspace $V[0]$ (the subspace killed by $\h$) of the tensor
product $V_1\otimes\cdots\otimes V_n$.  The arguments $z_1,\dots, z_n,\tau$
are complex numbers with $\tau$ in the upper half plane $H_+$, the $z_i$
are distinct modulo the lattice $\Z +\tau\Z$, and $\lambda\in\h$.
Introduce coordinates $\lambda=\Sigma\lambda_\nu h_\nu$ in terms
of an orthonormal basis ($h_\nu$) of $\h$. We use the notation
$X^{(i)}$ to denote the action of $X\in\End(V_i)$ on the
$i$th factor of a tensor product $V_1\otimes\cdots\otimes V_n$.
Similarly, if $X=\Sigma_l X_l\otimes Y_l\in\End(V_i)\otimes
\End(V_j)$, we set $X^{(ij)}=\sum_lX_l^{(i)}Y_l^{(j)}$.

In the formulation of
\cite{FW}, the KZB equations take the form
\begin{eqnarray}\label{KZB}
\kappa\partial_{z_j}u&=&
-\sum_\nu h_\nu^{(j)}\partial_{\lambda_\nu}u +\sum_{l:l\neq j}
\Omega_z^{(j,l)}(z_j-z_l,\tau,\lambda)u,\\
\kappa\partial_\tau u&=&
(4\pi i)^{-1}\triangle u
+\sum_{j,l}
\frac12\Omega_\tau^{(j,l)}(z_j-z_l,\tau,\lambda)u,
\end{eqnarray}
Here $\kappa$ is a complex non-zero parameter, $\triangle$ is
the Laplacian $\Sigma_\nu\partial_{\lambda_\nu}^2$
 and
\be\Omega(z,\tau,\lambda)=\Omega_z(z,\tau,\lambda)dz
+\Omega_\tau(z,\tau,\lambda) d\tau
\ee
is a differential form
with values in $\g\otimes\g$, with the following characterization.

For $l,m\in\Z$, let $S_{lm}$ be the transformation $(z,\tau)
\mapsto (z+l+m\tau,\tau)$ of $\C\times H_+$. Let
$L=\cup_{l,m}S_{lm}(\{0\}\times
H_+)$.  For any generic fixed $\lambda\in\h$, $\Omega$ is a
meromorphic differential 1-form
 on $\C\times H_+$ with values in $\g\otimes\g$ such that
{\rm (i)} $[x^{(1)}+x^{(2)},\Omega]=0$, for all $x\in\h.$
{\rm (ii)} $\Omega$ is holomorphic on $\C\times H_+-L$.
{\rm (iii)} $S_{lm}^*\Omega=\exp(2\pi i\ad^{(1)}_{\lambda})\Omega
-2\pi imd\tau C_0$.
{\rm (iv)} $\Omega$ has only simple poles and
$\Omega-Cdz/z$, is regular as $z\to 0$.

\begin{proposition}
For generic $\lambda$ there exist differential forms
obeying {\rm (i)-(iv)} and any two such forms differ by
a constant multiple of $C_0d\tau$. Moreover these forms
are closed and depend meromorphically on $\lambda\in\h$,
with simple poles on the hyperplanes in $\h$ defined
by the equations $\alpha(\lambda)
=l+m\tau$, $\alpha\in\Delta$, $l$, $m\in\Z$.
\end{proposition}
As stated in the proposition, the properties (i)-(iv) do
not characterize $\Omega$ completely. However, the KZB
equations are independent of the choice of $\Omega$
since $\Sigma_{ij} C_0^{(ij)}$ acts by zero on $V[0]$.
Explicitly, $\Omega$ has the form
\be
\Omega(z,\tau,\lambda)=\eta(z,\tau)C_0+\sum_{\alpha\in\Delta}
\omega_{\alpha(\lambda)}(z,\tau)C_\alpha,
\ee
The meromorphic differential forms $\eta$, $\omega_w$
on $\C\times H_+$ can be written in terms of
 Jacobi's theta function
\be
\vartheta_1(t,\tau)=-\sum_{j=-\infty}^{\infty}e^{\pi i(j+\frac12)^2\tau
+2\pi i(j+\frac12)(t+\frac12)},
\ee
as follows: introduce special functions (the prime denotes
derivative with respect to the first argument)
\be
\sigma_w(t,\tau)=\frac{\vartheta_1(w-t,\tau)\vartheta_1'(0,\tau)}
{\vartheta_1(w,\tau)\vartheta_1(t,\tau)},\qquad
\rho(t,\tau)=\frac{\vartheta_1'(t,\tau)}{\vartheta_1(t,\tau)}.
\ee
Then
\be
\omega_w(t)=\sigma_w(t,\tau)dt-
\frac1{2\pi i}\partial_w\sigma_w(t,\tau)d\tau,
\qquad
\eta=\rho(t,\tau)dt+\frac1{4\pi i}(\rho(t,\tau)^2+\rho'(t,\tau))
d\tau
\ee

\section{The KZB connection}\label{kzbc}
The compatibility of the system of equations \Ref{KZB}
can be expressed as the flatness of a connection. Consider
the action of $\Gamma=\Z\times\Z$ on $\C\times H_+$, defined
by $S_{l,m}$ above. Acting on each factor gives an action
of $\Gamma^n$ on $\C^n\times H_+$. Denote by $\pi:\C^n\times H_+
\to \C^n\times H_+/\Gamma^n$ the canonical projection onto
the space of orbits. Let $X_n=\C^n\times H_+/\Gamma^n-
\mbox{Diag}$, where Diag consists of orbits of
$(z_1,\dots,z_n,\tau)$ for which $z_i=z_j$ for some $i\neq j$.

Thus, for each representation $\rho$ of $\Gamma^n$ on a
vector space W, we get a vector bundle $B_\rho$ on $X_n$ which
is the restriction of $(\C^n\times H_+\times W)/\Gamma^n
\to(\C^n\times H_+)/\Gamma^n$.

In particular, we may take  $W=V[0]\otimes M(\h)$, where
$M(\h)$ is the space of meromorphic functions on $\h$,
and $\rho(\gamma)$, for $\gamma=((l_1,m_1),\dots, (l_n,m_n))$,
is multiplication by the End($V[0]$)-valued function
$\exp(-2\pi i\Sigma_jm_j\lambda^{(j)})$.

Thus, sections of $B_\rho$ over an open set $U\subset X_n$
are identified with functions $u(x,\lambda)$ on $\pi^{-1}(U)\times \h$
with values in $V[0]$, which are meromorphic on $\h$ for
all $x\in \pi^{-1}(U)$, and $\Gamma^n$-equivariant:
\be
u(\gamma\cdot x,\lambda)=\rho(\gamma)u(x,\lambda), \qquad (x,\lambda)
\in U\times\h,\quad \gamma\in\Gamma^n.
\ee
$B_\rho$ is a holomorphic vector bundle, if we declare that local
holomorphic sections are $V[0]$-valued $\Gamma^n$-equivariant
functions $u$  which are
meromorphic on $\pi^{-1}(U)\times\h$, and such that,
 for all $x\in\pi^{-1}(U)$,
$u(x,\cdot)$ is a meromorphic function on $\h$.

Here and below we use the notation $\Omega(z_i-z_j,\tau,\lambda)$
to denote the ($\lambda$-dependent)
differential form $p_{ij}^*\Omega$ on $X_n$ obtained by
pulling back $\Omega$ by the map $p_{ij}:(z,\tau)\to (z_i-z_j,\tau)$.
If $i=j$, $\Omega(0,\tau,\lambda)=\Omega_\tau(0,\tau,\lambda)d\tau$.

\begin{proposition}
The formula
\be
\nabla^{KZB}u=d u -\frac{d\tau}{4\pi i\kappa}\triangle u
+\frac1\kappa\sum_{i,\nu}dz_i h_\nu^{(i)}\partial_{\lambda_\nu}u
-\frac1{2\kappa}\sum_{i,j}\Omega^{(ij)}(z_i-z_j,\tau,\lambda)\,u
\ee
correctly defines a connection $\nabla^{KZB}: \Gamma(U,B_\rho)
\to\Gamma(U,B_\rho)\otimes\Omega^1(U)$, $U\subset X_n$.
This connection is flat and the KZB equations read
\be
\nabla^{KZB}u=0
\ee
\end{proposition}

\section{Weyl group and regularity}\label{wgr}
The coefficients of the KZB equations have singularities
 on the union of hyperplanes
\be
D=\cup_{\alpha\in\Delta}
\{(z,\tau,\lambda)\in X_n\times\h\,|\,\alpha(\lambda)\in\Z+\tau\Z\}.
\ee
Therefore a solution $u(z,\tau,\lambda)$ on $(U\times\h)-D$ will in
general be singular on $D$.

The Weyl group $W$ of $\g$ acts on
$V[0]$ and on $\h$, and thus on $V[0]$-valued functions
on $\h$. This action commutes with the representation $\rho$
of $\Gamma^n$, and thus  defines
 an action on the sections of $B_\rho$. It follows from the form
of the KZB equation that this action maps solutions to
solutions.
Let $\epsilon:W\to\{1,-1\}$ be the homomorphism mapping
reflections to $-1$. We say that a function $u:\h\to V[0]$ is
Weyl antiinvariant if $w\cdot u=\epsilon(w)u$ for all $w\in W$.
A section $u\in\Gamma(U,B_\rho)$ is Weyl antiinvariant if
it is Weyl antiinvariant as a function of $\lambda$ at each
point of $U$.

\begin{proposition}\label{reg}
Let $U$ be an open set in $X_n$ and
$u$ be a meromorphic Weyl antiinvariant solution of the KZB equation on
$U\times\h$, regular on $U\times \h-D$,
then $u$ extends to a holomorphic function on $U\times\h$.
Moreover, for all $\alpha\in\Delta$,  integers $r,s,l$, $l\geq 0$,
and $x\in\g_\alpha$,
\begin{equation}\label{van}
\bigl(\sum_{j=1}^ne^{2\pi isz_j}x^{(j)}\bigr)^l\,u
=O((\alpha(\lambda)-r-s\tau)^{l+1}),
\end{equation}
as $\alpha(\lambda)\to r+s\tau$.
\end{proposition}

\noi {\em Remark.} In the case where $\kappa$ is
an integer greater than or equal to the dual Coxeter number $h^\vee$ of
$\g$ and the highest weights obey $(\theta,\Lambda_j)\leq
\kappa-h^\vee$, $\theta$ being the highest root, the Weyl invariance and
the ``vanishing condition'' \Ref{van} appear as conditions for $u$
divided by the Weyl-Kac denominator  to be a conformal block of the
WZW model
\cite{FW} or to extend to an equivariant function
on the corresponding loop group \cite{EFK}. See Section \ref{ea}.

\section{The local system}\label{tls}
The first step in the construction of the integral representation is
the construction of a local system. Solutions will be expressed as
integrals over cycles with coefficients in this local system. Let $M$
be a positive integer, and define the family of configuration spaces
$X_M$ as above.  Fix $\mu=(\mu_1,\dots,\mu_M)\in\h^{*M}$, with
$\Sigma_i\mu_i=0$, and let $\chi_w$ be the character
$(l,m)\to\exp(-2\pi iwm)$ of $\Gamma=\Z\times\Z$.  Define a
representation of $\Gamma^M$ on the space $H(\h)$ of holomorphic
functions on $\h$: $\gamma=(\gamma_1,\dots,\gamma_M)$ acts by
multiplication by the function $\chi^\gamma_\mu:\lambda\mapsto
\chi_{\mu_1(\lambda)}(\gamma_1)\cdots\chi_{\mu_M(\lambda)}
(\gamma_M)$ on $\h$.  Let $B_{\mu}$ be the corresponding vector bundle
on $X_M$, constructed as above. Holomorphic sections on $U$ are viewed
as holomorphic $\Gamma^M$-equivariant functions on
$\pi^{-1}(U)\times\h$.
Let $E(t,\tau)=\vartheta_1(t,\tau)/\vartheta'_1(0,\tau)$.
Let $\Phi_\mu$ be the many-valued function
\be
\Phi_\mu=\prod_{i<j}E(t_i-t_j,\tau)^{(\mu_i,\mu_j)/\kappa}.
\ee
Then $\Phi_\mu^{-1}d\Phi_\mu$ is a single valued meromorphic differential
form on $\C^{M}\times H_+$.

\begin{lemmadef}
The formula
\be
\nabla u=d u -\frac{d\tau}{4\pi i\kappa}\triangle u
+\frac1\kappa\sum_{i,\nu}dz_i \mu_i(h_\nu)\partial_{\lambda_\nu}u
-\Phi_\mu^{-1}d\Phi_\mu\,u
\ee
correctly defines a connection $\nabla: \Gamma(U,B_\mu)
\to\Gamma(U,B_\mu)\otimes\Omega^1(U)$, $U\subset X_n$.
This connection is flat.
\end{lemmadef}

Let ${\cal L}_\mu$ be the corresponding local system of horizontal
sections. It can be described explicitly as follows: to give a
horizontal section $u\in\cal L_\mu(U)$ on a sufficiently small connected
neighborhood $U$ of a point in $X_M$, it is sufficient to give it as a
function on any lift $\tilde U$, a connected component of $\pi^{-1}(U)$.

\begin{proposition}
Let $U$ be a sufficiently small connected open neighborhood
of any point of $X_M$, and $\tilde U\subset \C^M\times H_+$
a lift of $U$. Then $\cal L_\mu(U)$ consists of $\Gamma^M$-equivariant
functions
on $\pi^{-1}(U)\times\h$ whose restriction to $\tilde U\times\h$
has the form
\be
\Phi_\mu(t_1,\dots,t_M,\tau)
g(\lambda-\kappa^{-1}\Sigma\mu_it_i,\tau),
\ee
for some choice of branch of $\Phi_\mu$ and some
holomorphic solution $g$ of the heat equation
\be
4\pi i\kappa\frac{\partial}{\partial\tau}g(\lambda,\tau)
=\triangle g(\lambda,\tau),
\ee
on $\h\times($projection of $U$ to $H_+)$.
\end{proposition}
If $\kappa$ is a positive integer, the vector bundle
$B_\mu$ has an interesting
finite rank subbundle  $\Theta^\kappa_\mu$ of ``theta functions'':
The fiber over $(t,\tau)\in X_M$ is the space of
holomorphic functions $u(\lambda)$ which are periodic
with respect to the coroot lattice $Q^\vee$ and
obey the relation
\be
u(\lambda+q\tau)=e^{-\pi i\kappa(q,q)\tau-2\pi i\kappa(q,\lambda)
+2\pi i\Sigma_j\mu_j(q)t_j}u(\lambda),\qquad \forall q\in Q^\vee.
\ee
Let $P$ be the weight lattice of $\g$.

\begin{lemma}\label{theta} The connection $\nabla$
preserves $\Theta^\kappa_\mu$. A basis of the space of horizontal
sections over $U\subset X_M$ is given by (branches of)
$\Phi_\mu(t,\tau)
\theta_{\kappa,p}(\lambda-\kappa^{-1}\Sigma_j\mu_j t_j,\tau)$,
where $\theta_{\kappa,p}$ are theta functions of level $\kappa$:
\be
\theta_{\kappa,p}(\lambda,\tau)
=
\sum_{q\in Q^\vee+p/\kappa}
e^{\pi i\kappa(q,q)\tau+2\pi i\kappa(q,\lambda)},
\ee
and $p$ runs over $P/\kappa Q^\vee$.
\end{lemma}

\section{Integral representation}\label{tir}
Let us begin by setting up the combinatorial framework
of our formula. It is essentially the same as in \cite{SV}.
We denote by $|A|$ the number of elements of a set $A$,
and by $S_n$ the group of permutations of $\{1,\dots,n\}$.
Choose a set $f_1,\dots,f_r$, $e_1,\dots,e_r$ of
Chevalley generators of $\g$ associated with simple
roots $\alpha_1,\dots,\alpha_r$. Fix highest weights
$\Lambda_1,\dots,\Lambda_n$ and let, as above, $V[0]$
be the zero weight space of the tensor product of the
corresponding irreducible
$\g$-modules, which is assumed to be non-trivial.
 We then have the decomposition
 $\Lambda_1+\cdots+\Lambda_n=\sum_jm_j\alpha_j$ with non-negative
integers $m_j$. Set $m=\Sigma_jm_j$. To each such sequence
of non-negative integers we can uniquely associate a ``color''
function $c$ on $\{1,\dots,m\}$
which is the only non-decreasing function
$\{1,\dots,m\}\to\{1,\dots,r\}$ such that $|c^{-1}(\{j\})|=m_j$
for all $1\leq j\leq n$.
Let $P(c,n)$  be the set of sequences
$I=(i_1^1,\dots,i_{s_1}^1;\dots ;i_1^n,\dots, i_{s_n}^n)$ of
integers in $\{1,\dots,r\}$, with $s_j\geq 0$, $j=1,\dots,n$ and
 such that, for all $1\leq j\leq r$,
$j$ appears precisely $|c^{-1}(j)|$ times in $I$.
For $I\in P(c,n)$, and a permutation $\sigma\in S_m$ set
$\sigma_1(l)=\sigma(l)$ and $\sigma_j(l)=\sigma(s_1+\dots+s_{j-1}+l)$,
$j=2,\dots,n$, $1\leq l\leq s_j$, and define $S(I)$ to be the
subset of $S_m$ consisting of permutations $\sigma$ such that
$c(\sigma_j(l))=i^j_l$ for all $j$ and $l$.

 Fix a highest weight vector $v_j$ for each representation
$V_j$. To every $I\in P(c,n)$ we associate a vector
\be
f_Iv=f_{i^1_1}\cdots f_{i^1_{s_1}}v_1\otimes
\dots\otimes f_{i^n_1}\cdots f_{i^n_{s_n}}v_n
\ee
in $V[0]$, and  meromorphic differential $m$-forms $\omega_{I,\sigma}$,
labeled by $\sigma\in S(I)$,
that we now define.

Let $\pi_j:\C^s\times H_+\to\C\times H_+$ be the projection
$(u_1,\dots,u_n,\tau)\mapsto (u_j,\tau)$. For $\lambda\in\h-D$
and $i_1,\dots,i_s\in\{1,\dots,r\}$, we define a  differential
$s$-form on $\C^s\times H_+$
\be
\omega_{i_1,\dots,i_s}(\lambda)=\pi_1^*\omega_{\alpha_{i_1}(\lambda)}
\wedge
\pi_2^*\omega_{(\alpha_{i_1}+\alpha_{i_2})(\lambda)}\wedge\cdots
\wedge\pi_s^*\omega_{\Sigma_{l=1}^{s}\alpha_{i_l}(\lambda)}.
\ee
Finally, for any pair $I\in P(c,n)$, $\sigma\in S(I)$, we have
$n$ maps $p_j:\C^{m+n}\times H_+\to\C^{s_j}\times H_+$,
defined by
\be
p_j(t_1,\dots,t_m,z_1,\dots,z_n,\tau)
=
(t_{\sigma_j(1)}-t_{\sigma_j(2)},t_{\sigma_j(2)}-t_{\sigma_j(3)},
\dots,t_{\sigma_j(s_j)}-z_j),
\ee
and a differential form
\be
\omega_{I,\sigma}(\lambda)=p_1^*\omega_{i_1^1,\dots,i^1_{s_1}}(\lambda)
\wedge\cdots\wedge
p_n^*\omega_{i_1^n,\dots,i^n_{s_n}}(\lambda).
\ee
The integrand of the integral representation of solutions
is  the $V[0]$-valued differential form
\be
\omega(\lambda)=\sum_{I\in P(c)}\sum_{\sigma\in S(I)}\sign(\sigma)
\omega_{I,\sigma}(\lambda)f_Iv.
\ee
\noindent{\it Examples:} (a) $\g=sl_3$, $n=1$, $\Lambda_1=\alpha_1+\alpha_2$
\be
\omega=\omega_{\alpha_1(\lambda)}(t_1-t_2)
\omega_{(\alpha_1+\alpha_2)(\lambda)}(t_2-z_1)\,f_1f_2v_1
-\omega_{\alpha_2(\lambda)}(t_2-t_1)
\omega_{(\alpha_1+\alpha_2)(\lambda)}(t_1-z_1)\,f_2f_1v_1.
\ee
(b) $\g=sl_2$, $n=1$, $\Lambda_1=2\alpha_1$.
\be
\omega=(\omega_{\alpha_1(\lambda)}(t_1-t_2)
\omega_{2\alpha_1(\lambda)}(t_2-z_1)
-\omega_{\alpha_1(\lambda)}(t_2-t_1)
\omega_{2\alpha_1(\lambda)}(t_1-z_1))\,f^2_1v_1.
\ee
(c) $\g=sl_2$, $n=2$, $\Lambda_1=\Lambda_2=\alpha_1/2$.
\be
\omega=\omega_{\alpha_1(\lambda)}(t_1-z_1)\,f_1v_1\otimes v_2
+\omega_{\alpha_1(\lambda)}(t_1-z_2)\,v_1\otimes f_1v_2.
\ee
Let $M=n+m$, and $\mu=(-\alpha_{c(1)},\dots,-\alpha_{c(m)},
\Lambda_1,\dots,\Lambda_n)\in\h^{*M}$, and let $t_1,
\dots,t_m$, $z_1,\dots,z_n,\tau$ be coordinates on $\C^M\times\h$,
with the action of $\Gamma^M$ as in the previous Section.
\begin{proposition}\label{om}
Let $\rho$ be as in Section 2. Then, for all $\gamma=(\gamma_1',
\dots,\gamma_m',\gamma_1'',\dots,\gamma_n'')\in\Gamma^M$,
\be
\gamma^*\omega(\lambda)=\chi_\mu^\gamma(\lambda)^{-1}
\rho(\gamma'')\omega(\lambda).
\ee
\end{proposition}
Let $p:X_M\to X_n$ be the projection onto the last $n+1$
factors.
It follows from Proposition \ref{om}
that $\omega$ can be viewed as a holomorphic
differential form on $X_M$ with values in Hom($B_\mu,p^*B_\rho$):
 if $s$ is any section of $B_\mu$,
then $\omega s$ is a differential form with values in $p^*B_\rho$.

Let us turn to the problem of integration of such differential
forms. The  homology of $X_M$ with coefficients in ${\cal L}_\mu$
can be computed with the complex $S_\cdot(X_M,{\cal L}_\mu)$
of singular chains. A $j$-chain is a linear combination
$\sum_\sigma s_\sigma\sigma$  of smooth $j$-simplices
$\sigma:\Delta_j\to X_M$ with coefficients
$s_\sigma\in\sigma^*{\cal L}_\mu(\Delta_j)$. This means that $s_\sigma$
maps a point $t$ in an affine $j$-simplex $\Delta_j$ to $s_\sigma(t)$
in
the stalk ${\cal L}_\mu(\sigma(t))$, so that for each $t\in\Delta_j$
there exists an open set $V\ni \sigma(t)$ and a section $\tilde s
\in{\cal L}_\mu(V)$ with $\tilde s\circ\sigma=s_\sigma$ on $\sigma^{-1}(V)$.
The boundary map is defined as usual on simplices and by restriction
of the sections to the boundary.

For each $x\in X_n$, the homology of the fiber $p^{-1}(x)$ is
the homology of the subcomplex of vertical chains $\sum s_\sigma\sigma$
with $p\circ\sigma(\Delta_j)=\{x\}$ for all $\sigma$ appearing
with non-zero coefficient.  A horizontal family of $j$-cycles on
an open $U\subset X_n$ is a linear combination of smooth maps
$\sigma:U\times\Delta_j\to X_M$ with coefficients $s_\sigma\in
\sigma^*{\cal L}_\mu(U\times\Delta_j)$, such that, for each $x\in U$,
$\sum s_\sigma(x,\cdot)\sigma(x,\cdot)$ is a vertical $j$-cycle on
the fiber $p^{-1}(x)$.

 If $\alpha$ is a differential $j$-form on $p^{-1}(U)$
with values in  Hom($B_\mu,p^*B_\rho$), and $\gamma$
is a vertical $j$-cycle on $p^{-1}(x)$ with coefficients in ${\cal L}_\mu$,
we may integrate $\alpha$ along $\gamma=\sum s_\sigma\sigma$:
\be
\int_\gamma\alpha:=\sum_{\sigma}\int_\sigma\alpha s_\sigma.
\ee
If the restriction of $\alpha$ to $p^{-1}(x)$ is closed with
respect to $\nabla$ (e.g., if $\alpha$ is a holomorphic
$m$-form), then the integral is independent of the representative
$\gamma$ in the homology class.

More generally, if $\gamma(x)$, $x\in U$ is a horizontal family of
$j$-cycles,
we may integrate $\alpha$ along each $\gamma(x)$ to get a
section $\int_\gamma\alpha$ of $B_\rho$.

Our main result is:
\begin{thm}
Let $U\subset X_n$ be open, and $\gamma(z,\tau)$, $(z,\tau)\in U$,
a horizontal family of $m$-cycles with coefficients in ${\cal L}_\mu$. Then
the section
\be
u(z,\tau)=\int_{\gamma(z,\tau)}\omega
\ee
of $B_\rho$ on $U$
is a solution of the KZB equation.
\end{thm}
This theorem follows from the following  technical key
result.

\begin{proposition}\label{key} Let $\mu$ be as above.
The  $V[0]$-valued differential form $\omega(\lambda)$
on $\C^M\times H_+$, with coordinates $t_1,\dots,t_m$,
$z_1,\dots,z_n,\tau$,  is
closed, and
\be
\kappa \Phi_\mu^{-1}d\Phi_\mu\wedge\omega(\lambda)
=\frac{d\tau}{4\pi i}
\wedge
\triangle\omega(\lambda)-
\sum_{i,\nu}dz_i\wedge h_\nu^{(i)}\partial_{\lambda_\nu}\omega(\lambda)
+\frac12\sum_{i,j}\Omega^{(ij)}(z_i-z_j,\tau,\lambda)
\wedge\omega(\lambda).
\ee
\end{proposition}

\section{Examples, applications}\label{ea}

\noindent{\bf A. Conformal field theory.}
The case of interest for conformal field theory is the case
where $\kappa=k+h^\vee$ with nonnegative integer level $k$
and $\Lambda_1,\dots,\Lambda_n$ obey the integrability
condition $(\Lambda_i,\theta)\leq k$.
The space of conformal blocks on an elliptic curve $\C/\Z+\tau\Z$
with $n$ points $z_1,\dots z_n$
with weights $\Lambda_1,\dots,\Lambda_n$
can be identified with a space  $E_\kappa(V,z,\tau)$
of theta functions with values in the zero weight
space $V[0]$ of $V=V_1\otimes\cdots\otimes V_n$ (see \cite{FW}).
By definition, $E_\kappa(V,z,\tau)$
is the space of holomorphic functions $u:\h\to V[0]$
with the properties that (i) $u$ is
 periodic with respect to the coroot lattice
$Q^\vee$ and
\be
u(\lambda+q\tau)=e^{-\pi i\kappa(q,q)\tau-2\pi i\kappa(q,\lambda)
+2\pi i\Sigma_jq^{(j)}z_j}u(\lambda),\qquad \forall q\in Q^\vee,
\ee
(ii) $u$ is $W$-antiinvariant, and (iii) $u$ obeys the
vanishing conditions \Ref{van}.

The space $E_\kappa(V,z,\tau)$ is the fiber of a holomorphic
vector bundle of finite rank over $X_n$ which is preserved
by the KZB connection  \cite{EFK}, \cite{FW}.
\begin{thm}\label{cft}
Suppose $\kappa=k+h^\vee$, $k=0,1,2,\dots$,
and $(\Lambda_j,\theta)\leq k$, $j=1,\dots,n$.
Let $\sum_{i=1}^{n}\Lambda_i=\sum_jm_j\alpha_j$,
with $\Sigma m_j=m$, and let $\mu$ be defined as in
Section \ref{tir}.
Then, for any horizontal family of $m$-cycles $\gamma(z,\tau)$,
with coefficients in the sheaf of horizontal sections
of $\Theta^\kappa_\mu$ (see Proposition \ref{theta})
the solution $\sum_{w\in W}\epsilon(w)w\cdot u(z,\tau,\cdot)$,
with $u(z,\tau,\lambda)=\int_{\gamma(z,\tau)}\omega$
 belongs to $E_\kappa(V,z,\tau)$.
\end{thm}

\medskip
\noindent{\bf B. Asymptotic solutions and eigenfunctions
of quantum $N$-body
systems.}
As explained in \cite{RV}, integral representations of
solutions of the Kni\-zhnik--Za\-mo\-lo\-dchi\-kov equation can
be used to construct common eigenvectors of the commuting systems
of operators appearing on the right-hand sides
of the equations, by applying the stationary phase method
to the integral.

The same procedure can be used here. The most interesting
case is when $n=1$. Then solutions are functions of
$\lambda\in\h$ and $\tau\in H_+$, with values in the zero
weight space of a
representation $V_1=V$. The KZB equations reduce to
\be
4\pi i\kappa\partial_\tau u=
(\triangle + \sum_{\alpha\in\Delta}\rho'(\alpha(\lambda),\tau)
e_\alpha e_{-\alpha})u
\ee
where $e_\alpha$ is a basis of $\g_\alpha$ with $(e_\alpha,e_{-\alpha})=1$.
The differential operator on the right-hand side is the Hamiltonian
of the so-called quantum elliptic Euler--Calogero--Moser model
\cite{ECM}, (for $sl_N$). This operator is part of a system
of $N-1$ commuting differential operators, whose symbols
are elementary symmetric functions \cite{EFK}. Note that, in terms
of Weierstrass' $\wp$ function with periods 1, $\tau$,
 $\rho'=-\wp+\eta_1$
for some  $\eta_1(\tau)$.

Let us describe explicitly the eigenvectors in a special case,
first considered by Etingof and Kirillov \cite{EK}, \cite{EK2} in which
the equation reduces to a scalar equation.  We take $\g=sl_N$
with $\h=\C^N/\C(1,\dots,1)$
and $V=S^{pN}\C^N$, the symmetric power of the defining
representation $\C^N$. Thus $V$ can be realized as the
space of homogeneous polynomials of degree $Np$ in
$N$ unknowns $x_1,\dots,x_N$.The weight zero
space $V[0]$ is one dimensional, spanned by $(x_1\cdots x_N)^p$ and,
for all roots $\alpha$, $e_\alpha e_{-\alpha}$ acts as
$p(p+1)$ on $V[0]$. The Weyl group $S_N$ acts on $V[0]$ trivially
if $p$ is even, and by the alternating representation if
$p$ is odd. The KZB equation is $4\pi i\kappa\partial_\tau u
=-H_{N,p} u$, and $H_{N,p}$ is the Hamilton operator of the elliptic
Calogero--Moser quantum N-body system \cite{OP}
\begin{equation}\label{CM}
-H_{N,p}=\sum_{i=1}^{N}\frac
{\partial^2}
{\partial\lambda_i^2}+2p(p+1)\sum_{i<j}\rho'(\lambda_i-\lambda_j,\tau),
\end{equation}
with coupling constant $p(p+1)$.

The highest weight of $V$ is $\Lambda=\sum_{j=1}^{N-1}(N-j)\alpha_j$.
The relevant color function in this case is the non-decreasing
function
\be c:\{1,\dots,m\}\to\{1,\dots,N-1\},\qquad m=N(N-1)p/2,\ee
 with $|c^{-1}\{j\}|
=(N-j)p$, $j=1,\dots,N-1$. The Chevalley generator $f_j$
corresponding to the simple
root $\alpha_j(\lambda)=\lambda_j-\lambda_{j+1}$
is represented by the differential
operator $f_j=x_{j+1}\partial/\partial x_{j}$ on $V\subset\C[x_1,
\dots, x_N]$.
If $s\in S_{m}$, introduce
a nonnegative integer $l(s)$ by
\be
f_{c(s(1))}\dots f_{c(s(m))}x_1^{Np}=l(s)(x_1\cdots x_{N})^p.
\ee
\begin{thm}\label{lame}
 Let $\xi\in\C^N$.
Suppose that $t\in\C^{m}$ obeys the ``Bethe ansatz'' equations
\be
\sum_{l:|c(l)-c(j)|=1}
\rho(t_j-t_l,\tau)
-
2
\sum_{l:l\neq j,c(l)=c(j)}
\rho(t_j-t_l,\tau)
+Np\delta_{c(j),1}\rho(t_j,\tau)=2\pi i\alpha_{c(j)}(\xi)
\ee
Then the function
\be
\psi(\lambda)=
e^{2\pi i\sum_{j=1}^N\xi_j\lambda_j}
\sum_{s\in S_{m}}l(s)
\prod_{j=1}^{m}
\sigma_{\sum_{l=1}^j\alpha_{c(s(l))}(\lambda)}(t_{s(j)}-t_{s(j+1)}),
\ee
with $t_{s(m+1)}:=0$,
is a meromorphic eigenfunction of $H_{N,p}$ with eigenvalue
\bea
\epsilon&=&4\pi^2\sum_j\xi_j^2-4\pi i\partial_\tau S(t_1,\dots,t_m,\tau), \\
S(t_1,\dots,t_m,\tau)&=&\sum_{i<j}(2\delta_{c(i),c(j)}\ln E(t_i-t_j,\tau)
-\delta_{|c(i)-c(j)|,1}\ln E(t_i-t_j,\tau))\\
& & -Np\sum_{c(i)=1}\ln E(t_i,\tau).
\eea
Moreover, $\psi$ is regular off the root hyperplanes
$\lambda_i=\lambda_j$, $i<j$.
\end{thm}
This theorem follows by computing the asymptotics of the
integral representation when $\kappa$ goes to zero
(with a non-degeneracy assumption), or, more directly,
from Proposition \ref{key}. The regularity off root
hyperplanes follows from the regularity of the differential
equation. It then follows from Proposition \ref{reg}
that the Weyl averaged eigenfunction
\be
\psi^W(\lambda)=\sum_{w\in S_N}\epsilon(w)^{p+1}\psi(w\cdot\lambda)
\ee
is holomorphic on all of $\C^N$.

In the case $N=2$, Theorem \ref{lame}
reduces to Hermite's 1872 solution
of the Lam\'e equation (see \cite{WW}, 23$\cdot$71), and was rederived
in \cite{EK} using the asymptotics of the integral solutions
of the KZB equation for $sl_2$ \cite{BF}.

\medskip
\noindent{\bf C. Integral identities.}
Consider again the case $\g=sl_N$, $n=1$, $V=S^{Np}\C^N$,
but now for integer $\kappa\geq N$ as in A. The representation
$V$ obeys the integrability condition if $N(p+1)\leq\kappa$.
The limiting case $\kappa=N(p+1)$ is particularly interesting
as the space of conformal blocks is one dimensional and can be
described explicitly:

\begin{proposition} \cite{EK2} Let $\g=sl_N$, $n=1$, $V=S^{Np}\C^N$, and
$\kappa=N(p+1)$.  Then the space $E_\kappa(V,0,\tau)$ is
one-dimensional, and is spanned by the $(p+1)$th power of the
Weyl--Kac denominator
 \be
\Pi(\lambda,\tau) =q^{\frac{(N^2-1)}{24}}
\prod_{j<l}(e^{i\pi(\lambda_j-\lambda_l)}
-e^{i\pi(\lambda_l-\lambda_j)})\prod_{m=1}^\infty
\left[(1-q^m)^{N-1}\prod_{j\neq l}(1-q^me^{2\pi
i(\lambda_j-\lambda_l)})\right],
\ee
$q=\exp(2\pi i\tau)$. Moreover, $\Pi^{p+1}$ is a horizontal section
for the KZB connection:
\be 4\pi iN(p+1)\partial_\tau
\Pi(\lambda,\tau)^{p+1}=-H_{N,p}\Pi(\lambda,\tau)^{p+1},
\ee
where $H_{N,p}$ is the differential operator $($\ref{CM}$)$.
\end{proposition}
It follows that the integrals of Theorem \ref{cft} in
the case considered here are proportional to $\Pi^{p+1}$, and
the proportionality factor can be computed in the
limit $\tau\to i\infty$ in terms of Selberg type integrals
leading to non-trivial integral relations involving theta
functions. Let us give here the simplest one, obtained
in the case $N=2$, $p=1$.

\begin{thm}\label{intid}
Let $\theta_{4,m}(x,\tau)=\sum_{j\in\Z}e^{\pi i(8j+m)^2\tau/8
+\pi i(8j+m)x}$, $m\in\Z/8\Z$. Then the integral
\be
h_m(x,\tau,\kappa)=
\int_0^1E(t,\tau)^{-\frac 2\kappa}
[\sigma_x(t)\theta_{4,m}(x+2t/\kappa)+
\sigma_{-x}(t)\theta_{4,m}(-x+2t/\kappa)]dt
\ee
converges for Re$(\kappa)<0$, and
has an analytic continuation to a meromorphic function
regular at $\kappa=4$, and  $h_m(x,\tau,4)$
vanishes identically at $\kappa=4$
unless $m\equiv 2\mod 4$. If the branch of the logarithm
is chosen in such a way that $\arg(E(t,\tau))\to 0$ as
$t\to 0^+$, then if  $m\equiv 2\mod 4$,
\be h_m(x,\tau,4)=2\pi^{1/2}B(-{\scriptstyle{\frac12},{\frac34}})
[q^{1/8}
(e^{i\pi x}-e^{-i\pi x})
\prod_{j=1}^\infty
(1-q^j)(1-q^je^{2\pi ix})(1-q^je^{-2\pi ix})]^2,
\ee
where $B$ is Euler's beta function and $q=\exp(2\pi i\tau)$.
\end{thm}
This identity is similar to the identity given
in \cite{F}, which was based
on the (conjectural) identification of the (3,4) minimal Virasoro model
with the scaling limit of the Ising model.

\end{document}